\documentstyle[aps,preprint]{revtex}
\tighten
\preprint{IHEP-TH-25}
\begin{document}
\draft
\title{Light-cone QCD predictions for elastic $ed$-scattering 
in the intermediate energy region}
\author{Jun Cao$^{2}$ and  Hui-fang Wu$^{1,2}$}
\address{
{1. China Center of Advanced Science and Technology 
(World Laboratory)} \\
{P.O. Box 8730, Beijing, 100080, China}\\
{2. Institute of High Energy Physics, P.O. Box 918(4), Beijing,
100039, China\thanks{Mailing address.}  }
}
\date{\today}
\maketitle
\begin{abstract}
The contributions of helicity-flip matrix elements to the
deuteron form factors are discussed in the light-cone frame.
Normalized $A(Q^2)$, $B(Q^2)$, $G_Q(Q^2)$ and $T_{20}$ are obtained in
a simple QCD-inspired model. We find that $G_{+-}^+$ plays an important
role in $G_Q(Q^2)$.
Our numerical results are consistent with the data in the intermediate 
energy region. 
\end{abstract}
\narrowtext

\section{INTRODUCTION}

The electromagnetic properties of the deuteron have received extensive
attention since it can be used to explore the quark and gluon degrees of
freedom in the simplest nuclei. Many approaches are suggested to explain 
its electromagnetic form factors. Among them, the conventional
meson-nucleon picture \cite{ga-hy,cckp} gives a satisfying explanation in
the low energy region, i.e. $Q^2 < 1$ GeV$^2$. However, as mentioned by
Arnold {\it et al.} \cite{arnold}, the experimental results are in sharp
disagreement with the meson exchange calculations at higher momentum
transfer (see Fig.~1). It means that quark and gluon degrees of freedom must
be taken in account. Moreover, the success of dimensional counting rules
\cite{br-ch} implies that perturbative QCD(pQCD) may give correct
predictions at large momentum transfers. 
PQCD predicts~\cite{2157} that, in the Light-cone frame(LCF), the
helicity-zero to zero matrix element $G^+_{00}$ 
will be the dominant helicity amplitude at large $Q^2$ for elastic 
{\it ed}-scattering. Carlson and Gross \cite{ca-gr} have pointed out that LCF
helicity-flip amplitudes, $G^+_{+0}$ and $G^+_{+-}$, are suppressed by
factors of $\Lambda_{\rm QCD}/Q$ and $\Lambda_{\rm QCD}^2/Q^2$,
respectively. It is argued \cite{bh92} that the dominance of $G^+_{00}$,
thus the validity of perturbative QCD predictions, begins at $Q^2 \sim 0.8$
GeV$^2$. Assuming the helicity-zero to zero dominance, A simple ratio of
form factors for the deuteron is predicted
\begin{equation}
G_C : G_M : G_Q = (1-\frac{2}{3}\eta) : 2 : -1 ,
\end{equation}
with the kinematic factor $\eta = Q^2/4M^2$.

\par
Unfortunately, only to the lowest order, over 300 000 diagrams containing
six fermion lines connected by five gluons are required to obtain the full
elastic {\it ed}-scattering amplitude in pure pQCD approach. Such a
calculation can be found in  the work of Farrar, Huleihel, 
and Zhang \cite{fhz}, which shows that the direct application of pQCD
to deuteron form factor at experimentally accessible momentum transfers has
still a long way to go.
\par
To make detailed predictions for deuteron electromagnetic form factors, we
have suggested a model \cite{cw} in the intermediate energy region. Based on
the reduced nuclear amplitude defined by Brodsky and Hiller \cite{bh83}, the
model makes the points that a simple nuclear wave function can represent the
data of the deuteron electromagnetic structure function $A(Q^2)$ and shows a
scenario where $G^+_{00}$ is already dominant at $Q^2$ of 1 GeV$^2$.
Normalized $G_{00}^+$ can be extracted from fitting the data of $A(Q^2)$. It
is pointed out \cite{ko-sy} that the helicity-flip matrix element $G^+_{+0}$
can not be neglected in the expression of $B(Q^2)$. In addition to that, we
find that $G^+_{+-}$ plays an important role in $G_Q(Q^2)$. Neglecting it
will result in a contradiction with the data and the conventional
meson-nucleon prediction in the low energy region. Due to the kinematical
reason, the ratio of form factors in Eq.~(1) should be modified
even at extremely high $Q^2$.
We will model the helicity-flip matrix elements $G_{+0}^+$ and
$G_{+-}^+$ in the intermediate energy region according to the pQCD
predictions. The normalized structure functions $A(Q^2)$ and $B(Q^2)$,
tensor polarization $T_{20}$, and
form factors $G_C(Q^2)$ and $G_Q(Q^2)$ will be discussed on the basis of the
normalized $G_{00}^+$ and the above model.
\par
A QCD-inspired model for $G^+_{00}$ will be reviewed in Sec.~II. Under a
phenomenological consideration, the model is applied to obtain normalized
$A(Q^2)$, $B(Q^2)$ and $T_{20}$ in Sec.~III, 
where $G_Q$ is also discussed. The numerical results and summary are
presented in Sec.~IV and Sec.~V, respectively.

\section{ A QCD-INSPIRED MODEL}
\par
For electron scattering on a deuteron, the Rosenbluth cross section
is
\begin{equation}
\frac{d\sigma}{d\Omega}=\left( \frac{d\sigma}{d\Omega} \right)_{\rm Mott}
\left[ A(Q^2)+B(Q^2)\tan^2(\frac{\theta}{2}) \right]   ,
\end{equation}
where $A(Q^2)$ and $B(Q^2)$ are determined by \cite{acg}:
\begin{eqnarray}
A(Q^2) & = & G_C^2+\frac{2}{3} \eta G_M^2+\frac{8}{9}\eta^2 G_Q^2
   ,\\
B(Q^2) & = & \frac{4}{3}\eta (1+\eta )G_M^2    ,
\end{eqnarray}
with $\eta =Q^2/4M^2$, where $M$ is the deuteron mass.
The deuteron form factor $F_d(Q^2)$ is defined as 
$F_d(Q^2)\equiv\sqrt{A(Q^2)}$.

\par
In the standard LCF, defined by \cite{dr-ya} $q^+=0, q_y=0$, and $q_x=Q$,
the above form factors can be obtained from the plus component of three
helicity matrix elements:

\begin{mathletters}
\label{gcmq1}
\begin{eqnarray}
G_C & = &  \frac{1}{2p^+(2\eta+1)}\left[ (1-\frac{2}{3}\eta) G_{00}^+ +
\frac{8}{3}\sqrt{2\eta}  G_{+0}^+ +\frac{2}{3}(2\eta-1) G_{+-}^+ \right]
  ,\label{gc} \\
G_M & = &  \frac{1}{2p^+(2\eta+1)}\left[ 2 G_{00}^+ +\frac{2(2\eta-1)}
{\sqrt{2\eta}}  G_{+0}^{+} -2 G_{+-}^+ \right]
  , \label{gm}\\
G_Q & = &  \frac{1}{2p^+(2\eta+1)}\left[ - G_{00}^+ +\sqrt{\frac{2}{\eta}}
 G_{+0}^{+} -\frac{\eta+1}{\eta} G_{+-}^+ \right]
  . \label{gq}
\end{eqnarray}
\end{mathletters}
\par
To avoid the complicated pQCD calculation, a model for the deuteron form
factor $F_d(Q^2)$ has been suggested \cite{cw}. The point is that the hard
kernel at large $Q^2$ is assumed to be the perturbative amplitude for the
six quarks to scatter from collinear to the initial two-nucleon
configuration to collinear to the final two-nucleon configuration, where
each nucleon has roughly equal momentum. Since the gluon is a color octet,
the single-gluon exchange between two color-singlet nucleon is forbidden. In
the lowest order, the quark-interchange is necessary in addition to the
one-gluon exchange between two nucleons. For the binding energy of the
deuteron is small, we can divide roughly the kernel into two parts. One
represents the interchange of quarks and the gluon exchange between two
nucleons, which transfer about half of the transverse momentum of the
virtual photon from the struck nucleon to the spectator nucleon. Another
part is the inner evolution of two nucleons. The first part leads to the
reduced form factor of the deuteron and the latter leads to the form factors
of two nucleons. A vector boson(color singlet) with an effective mass $M_b$
is introduced to represent the quark-interchange and the one-gluon exchange
effects.
\par
Assuming the $G^+_{00}$ dominance in the structure function $A(Q^2)$,  
we get \cite{cw}
\begin{equation}\label{g00}
G^+_{00}(Q^2) = 2(2\eta +1) F_N^2(Q^2/4)
\int [dx][dy]\phi_d^{\dag}(x,Q) t_H(x,y,Q) \phi_d(y,Q), 
\end{equation}
where $t_H$ is the hard scattering amplitude and $\phi_d$ is the body
distribution amplitude of the deuteron defined by
\begin{equation}\label{da}
\phi_d(x,Q)=\int [d{\bf k}_{\perp}] \Psi_d^{\rm body}(x_i,{\bf k}_{\perp i}) 
  .
\end{equation}
The argument for the nucleon form factor, $F_N$, is $Q^2/4$ since, in 
the limit of zero binding energy, each nucleon must change its momentum
from $P/2$ to $(P+q)/2$.
Using Brodsky-Huang-Lepage prescription \cite{bhl} from a harmonic
oscillator wave function, $A^{\prime}\exp [-\frac{1}{2}\alpha^2 r^2]$, in the 
rest frame, the body wave function can be written as
\begin{equation} \label{wf}
\Psi_d^{\rm body}(y,{\bf l}_{\perp} )= 
A \exp\left[ - \frac{1}{ 2{\alpha}^2 } 
\frac{{\bf l}_{\perp}^2 + m_N^2 }{4 y (1-y} \right] ,
\end{equation}
where $A$ is determined by the normalization of the wave function
and $\alpha$ by fitting the data of $A(Q^2)$.
A direct calculation of diagrams in Fig.~2 gives
\begin{equation}\label{th}
t_H(x,y,Q)=\frac{4M^2 g_{\rm eff}^2}{xy Q^2 +{M_b}^2- { (x-y) }^2 M^2}
\frac{1}{xQ^2+(\frac{1}{4}- {(1-x)}^2 M^2) }  ,
\end{equation}
where $g_{\rm eff}$ is an effective coupling constant. 
We have taken the nucleon mass 
to be half of the deuteron mass in Eq.~(\ref{th}).

\section{THE ROLE OF HELICITY-FLIP MATRIX ELEMENTS}
\par
In the intermediate energy region, $G^+_{00}$ dominates the charge form
factor $G_C$, but not $G_M$ and $G_Q$, because the kinematic factor $\eta$
is still small. While $\eta \ll \frac{1}{2}$, the $G^+_{+0}$ contributions
to both $G_M$ and $G_Q$ are enhanced by a factor of $1/\sqrt{2\eta}$. The
$G^+_{+-}$ contribution to $G_Q$ is enhanced by $1/\eta$. Although
$G^+_{+0}$ and $G^+_{+-}$ are suppressed for dynamical reason, they may
contribute significantly to $G_M$ and $G_Q$ because of the kinematic
enhancement. Since $G_C$ dominate $A(Q^2)$ while $\eta$ is small, the
$G^+_{00}$ dominance works very well in determining $A(Q^2)$. As for
$B(Q^2)$, the helicity-flip matrix element $G^+_{+0}$ must be taken into
account \cite{ko-sy}. In $G_Q(Q^2)$, both $G^+_{+0}$ and $G^+_{+-}$ may be
important.

\par
As is well known, pQCD predicts that $G_{+0}^+$ and $G_{+-}^+$ are
suppressed by factors $\Lambda_{\rm QCD}/Q$ and $\Lambda_{\rm QCD}^2/Q^2$,
respectively. The QCD scale $\Lambda_{\rm QCD}$ is around 200 MeV. In order
to explore the role of $G^+_{+0}$ and $G^+_{+-}$, we assume the
following relations phenomenologically~:
\begin{mathletters}\label{gpm}
\begin{eqnarray}
G_{+0}^+ &=&  \frac{f}{\sqrt{2\eta}} G_{00}^+   ,\\
G_{+-}^+ &=& c\frac{f^2}{2\eta} G_{00}^+   ,
\end{eqnarray}
\end{mathletters}
where $f$ and $c$ are two parameters. These relations are expected to be
reasonable as $Q^2 \gtrsim 2 M \Lambda_{\rm QCD} \sim 0.8$ GeV$^2$, where 
$2M \Lambda_{\rm QCD}$ is a scale that determines the $G_{00}^+$ dominance
\cite{bh92}. Ref.~\cite{ko-sy} reveals that $G_{+0}^+$ contributes
significantly to $G_M(Q^2)$. The $G_{+0}^+$ contribution to $G_Q(Q^2)$ was
also discussed there. However, as mentioned above, $G_{+-}^+$ should be
taken into account in determining $G_Q$ due to the same reason as
$G_{+0}^+$. Detailed analysis shows that it plays an important role in the
intermediate energy region. By assuming relation (\ref{gpm}) we have picked
up the contributions of $G_{+0}^+$ and $G_{+-}^+$ which may be comparable
with that of $G_{00}^+$ to $G_M$ and $G_Q$ in the intermediate energy
region. Nonleading contributions to $G_{+0}^+$ that may enter at the same
order of $G_{+-}^+$ are neglected for kinematic reason. Inclusion of these
terms will change the magnitude of $f$ slightly.

\par
Substituting Eq.~(\ref{gpm}) into Eq.~(\ref{gq}), $G_Q$ becomes
\begin{equation}\label{gqc}
G_Q = \frac{1}{2 P^+ (2\eta+1) } \left[ -1 + \frac{f}{\eta} -
\frac{\eta+1}{\eta} \frac{cf^2}{2\eta} \right] G_{00}^+   .
\end{equation}
If $G_{+-}^+$ is suppressed strongly, $G_Q(Q^2)$ will be negative while
the pQCD begin to be valid. Since it is positive at the origin, there must
be a node in the region of $Q^2 < 1$ GeV$^2$. It is in sharp disagreement
with the meson-nucleon picture (e.g. see Ref.~\cite{ga-hy}) and the
experimental data \cite{garcon}. Another constraint on the parameters can be
obtained from the data of $G_M(Q^2)$, which reveals that
$G_M(Q^2)$ changes its sign at $Q^2 = Q_0^2 \sim 2 {\rm GeV}^2$. It turns
out 
\begin{equation}\label{gmc}
2 \eta_0 + (2 \eta_0 -1) f - c f^2 = 0,
\end{equation}
with $\eta_0 = Q_0^2 /4 M^2$. Combining Eq.~(\ref{gqc}) and Eq.~(\ref{gmc}), 
it is shown
that $G_Q$ will keep positive at any momentum transfers if $c > 0.43$.
Different choices of $c$ will
produce very different $G_Q$, but have little effect on $A(Q^2)$ and $B(Q^2)$.

\section{ NUMERICAL RESULTS}
\par
At first, we assume the $G^+_{00}$ dominance in $A(Q^2)$, from which
we get Eq.~(\ref{g00}). By fitting the data of $A(Q^2)$(see Fig.~1)
we obtain the parameters:
$M_b = 0.5 {\rm GeV}, \alpha =0.21 {\rm GeV}$ and
$\alpha_{\rm eff} = g^2_{\rm eff}/4\pi = 0.15$.
\par
Then,
given $c$, we can get the expressions of $G_C$, $G_M$, and $G_Q$ by
substituting Eq.~(\ref{gpm}) into Eqs.~(\ref{gcmq1}). 
As a demonstration, we will show the $c=1$ case for simplicity.
In this case $f$ is $2 \eta_0$.
\begin{mathletters}
\begin{eqnarray}
G_C & = &  \frac{G^+_{00}}{2(2\eta+1)} \left[ \frac{16}{3}\eta_0
+\frac{8}{3}\eta_0^2+1
-\frac{2}{3}\eta-\frac{4}{3}\frac{\eta_0^2}{\eta} \right]
  ,\label{gc2} \\
G_M & = &  \frac{G^+_{00}}{2(2\eta+1)} \left[ 2(2\eta_0+1) 
(1-\frac{\eta_0}{\eta} ) \right]     , \label{gm2}\\
G_Q & = &  \frac{G^+_{00}}{2(2\eta+1)} \left[ 
-\frac{\eta_0^2}{\eta^2}(2\eta+1) -(1-\frac{\eta_0}{\eta})^2 \right]
  , \label{gq2}
\end{eqnarray}
\end{mathletters}
with $\eta_0 = Q_0^2/4 M^2$. The inclusion of helicity-flip matrix elements
has little effect on $A(Q^2)$ (see Fig.~1), but it changes the $\alpha_{\rm
eff}$ to 0.11. The normalized structure function $B(Q^2)$ is given in
Fig.~3, where we have chosen the parameter $Q_0^2 = 1.85$ GeV$^2$ (i.e.
$\eta_0 = 0.13$) from fitting the data. From Eq.~(\ref{gq2}) it is easy
seen that $G_Q$ keeps positive for all momentum transfers for $G_{00}^+$ is
negative. For different choices of $c$, the normalized $G_Q$ and $T_{20}$
are shown in Fig.~4 and Fig.~5, respectively. The corresponding magnitudes 
of $f$ are shown there, too. At very large momentum
transfers, say $\eta \gg \eta_0$, the ratio of form factors will be slightly
modified to
\begin{equation}
G_C: G_M: G_Q =( ( 1+\case{8}{3} f +\case{2}{3} c f^2 )
	-\case{2}{3}\eta ) : 2 (1+f): -1   .
\end{equation}	

\section{SUMMARY}
\par
We have discussed the electromagnetic form factors of the deuteron in a QCD
inspired model. Detailed kinematic analysis in LCF reveals that, provided
the validity of pQCD, the helicity-zero to zero matrix element $G_{00}^+$
dominates the gross structure function $A(Q^2)$ in both the large and
intermediate energy region, which is used to extract the normalized
$G_{00}^+$ from a simple model for $A(Q^2)$. Further analysis shows that
$G_{+0}^+$ and $G_{+-}^+$ are also important to determine other form
factors. In the present work, $G_{+0}^+$ and $G_{+-}^+$ are modeled
according to some pQCD predictions at high $Q^2$. Normalized $B(Q^2)$ is
obtained, whose vanishing at $Q_0^2 = 1.85$ GeV$^2$ is used to determine the
extent that helicity-flip matrix elements are suppressed to. To which extent
the $G_{+-}^+$ is suppressed will strongly effect the behavior of $G_Q$ in
the intermediate energy region. We find that, If $G_{+-}^+$ is suppressed
strongly, $G_Q$ will change its sign twice and one of its nodes lies in the
$Q^2 < 1$ GeV$^2$ region. It is contrary to the meson-nucleon picture and
experimental data. If $c > 0.43$, $G_Q$ will keep positive for all momentum
transfers in our model. Different choices of $c$ have little effect on
$A(Q^2)$ and $B(Q^2)$. As an example, we demonstrate a simple case by
choosing $c = 1$, but a larger $c$ shows a better fit for $G_Q$ and
$T_{20}$. Since the momentum transfers of available data are not high enough
to determine the value of $c$ reliably, we just show a qualitative result.
At very high $Q^2$, the ratio of form factors, $G_C : G_M : G_Q$, will be
slightly modified. The ratio of $G_C$ and $G_Q$ is the same as predicted by
pQCD as $Q^2 \rightarrow \infty$ \cite{ca-gr}. It is apparent that $G_M$ is
bigger than that predicted by Brodsky and Hiller \cite{bh92}. The kinematic
factor $\sqrt{2\eta} - 1/\sqrt{2\eta}$ leads to the contributions of
$G_{+0}^+$ to $G_M$ in both the large and intermediate energy regions.
\par
By assuming relation (\ref{gpm}) we have picked up the leading contributions
of $G_{+0}^+$ and $G_{+-}^+$ which may be comparable with that of $G_{00}^+$
to $G_M$ and $G_Q$ in the intermediate energy region. There are corrections
from the next to leading order contributions to $G_{+0}^+$ which have the
form similar to $G_{+-}^+$. For $G_Q$ these corrections can be neglected
because of the lack of kinematic enhancement as $G_{+-}^+$ has. For $G_M$
they enter at the same order as $G_{+-}^+$. Inclusion of these terms will
diminish the magnitude of parameter $f$ in the same way as increasing
parameter $c$. Since $G_{+-}^+$ is not important in $G_M$ itself(i.e. $f$ is
insensitive to $c$), our conclusions will not suffer from neglecting them.
Precise $f$ can not be determined with these terms unknown. It is shown in
Fig.~4 and Fig.~5 that $f$ is around 0.2.

\par
Since we obtain the normalized $G_{00}^+$ by assuming its dominance in
$A(Q^2)$ and relation (\ref{gpm}) by the validity of pQCD, our conclusions,
except the ratio, are valid only in the intermediate energy region. It is
found that $G_C$ has a node \cite{garcon} at $Q^2=0.75$ GeV$^2$, which is
predicted by conventional meson-nucleon method, too. Although our model can
not predict the accurate position, it still shows that the node lies at
somewhere $Q^2 < 1$ GeV$^2$. At very large $Q^2$, the contributions of
hidden-color states should be taken into account to get a normalized
$G_{00}^+$. To some extent, our model shows a smooth connection with pQCD
predictions in the high energy region and with traditional nuclear physics
conclusions in the low energy region. Thus it can be expected to unify the
predictions for the deuteron form factors from the low energy to large energy
region. The experiments at CEBAF will be crucial to build a more realistic
model in the intermediate energy region.

\acknowledgments

The authors would like to thank Professor T. Huang for his
valuable discussions.
\par
This work was supported by National Science Foundation of China (NSFC)
and Grant No. LWTZ-1298 of Academia Sinica.



\begin{figure}
\caption{Structure function $A(Q^2)$ of the elastic $ed$-scattering 
in our model(the solid line), with $M_b=0.5$ GeV and $\alpha = 0.21$ GeV,
The dashed line corresponds to the Paris potential calculation.
Experimental data are taken from Ref.~\protect\cite{arnold,platchkov}.
The dotted line is our result with the corrections from $G_{+0}^+$
and $G_{+-}^+$, which is almost overlapping with the solid line.}
\label{fig1} 
\end{figure}
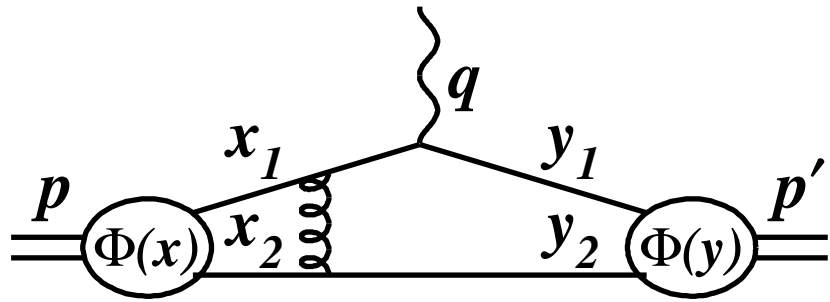
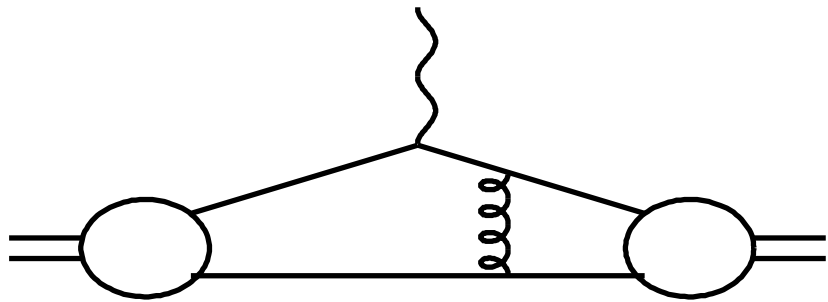
\begin{figure}
\caption{
The hard scattering diagrams.
}
\label{fig2}
\end{figure}
\begin{figure}
\caption{
Structure function $B(Q^2)$ (the solid line). 
The dashed line corresponds to the Paris potential calculation.
Experimental data are taken from Ref.~\protect\cite{bform}.
}
\label{fig3}
\end{figure}
\begin{figure}
\caption{
The form factor $G_{Q}$.
Experimental data are taken from Ref.~\protect\cite{garcon}.
}
\label{fig4}
\end{figure}
\begin{figure}
\caption{
The tensor polarization $T_{20}$ with scattering angle $\theta = 70^\circ$. 
The dashed line corresponds to the
calculation with Paris potential. 
}
\label{fig5}
\end{figure}
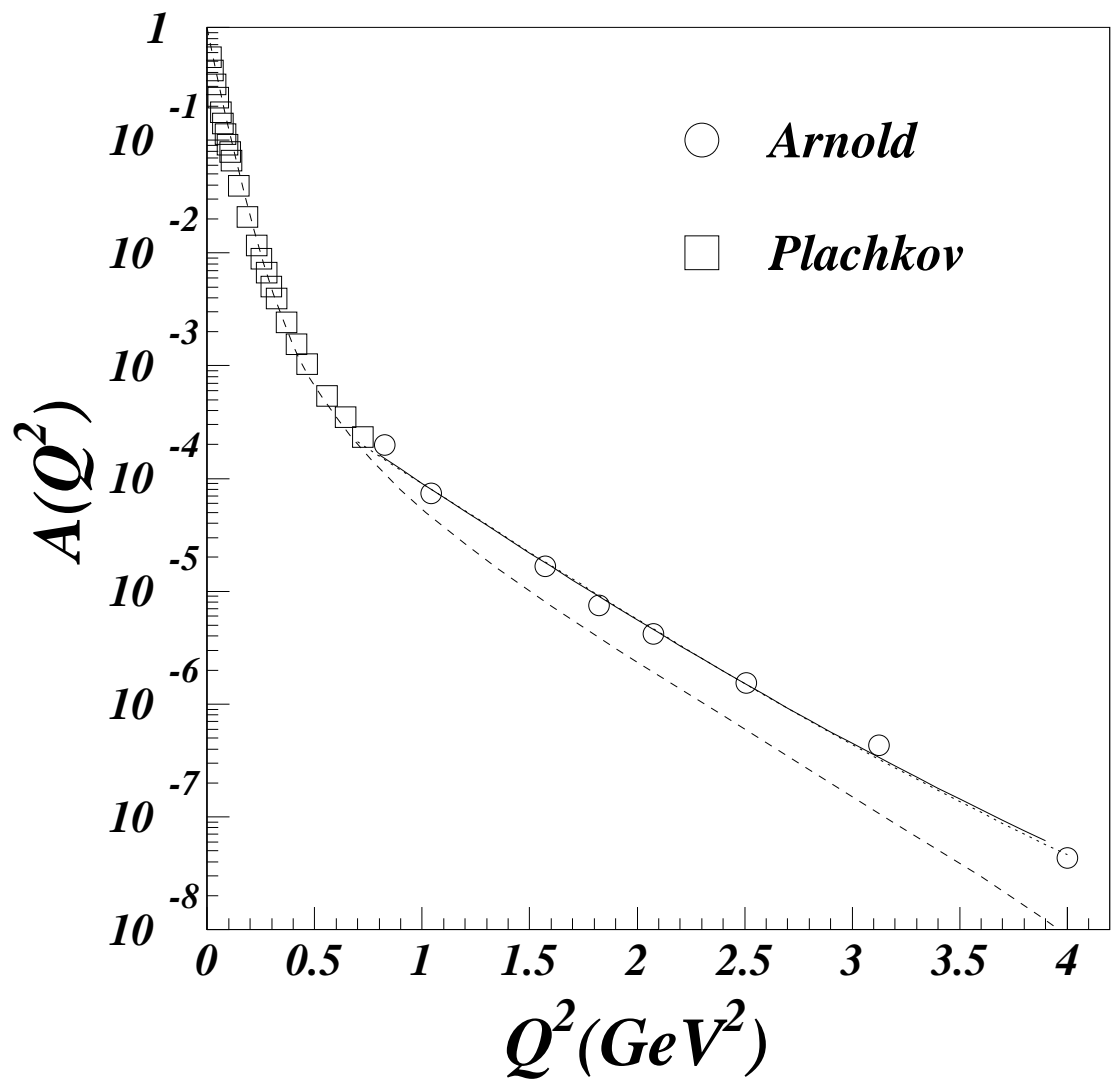
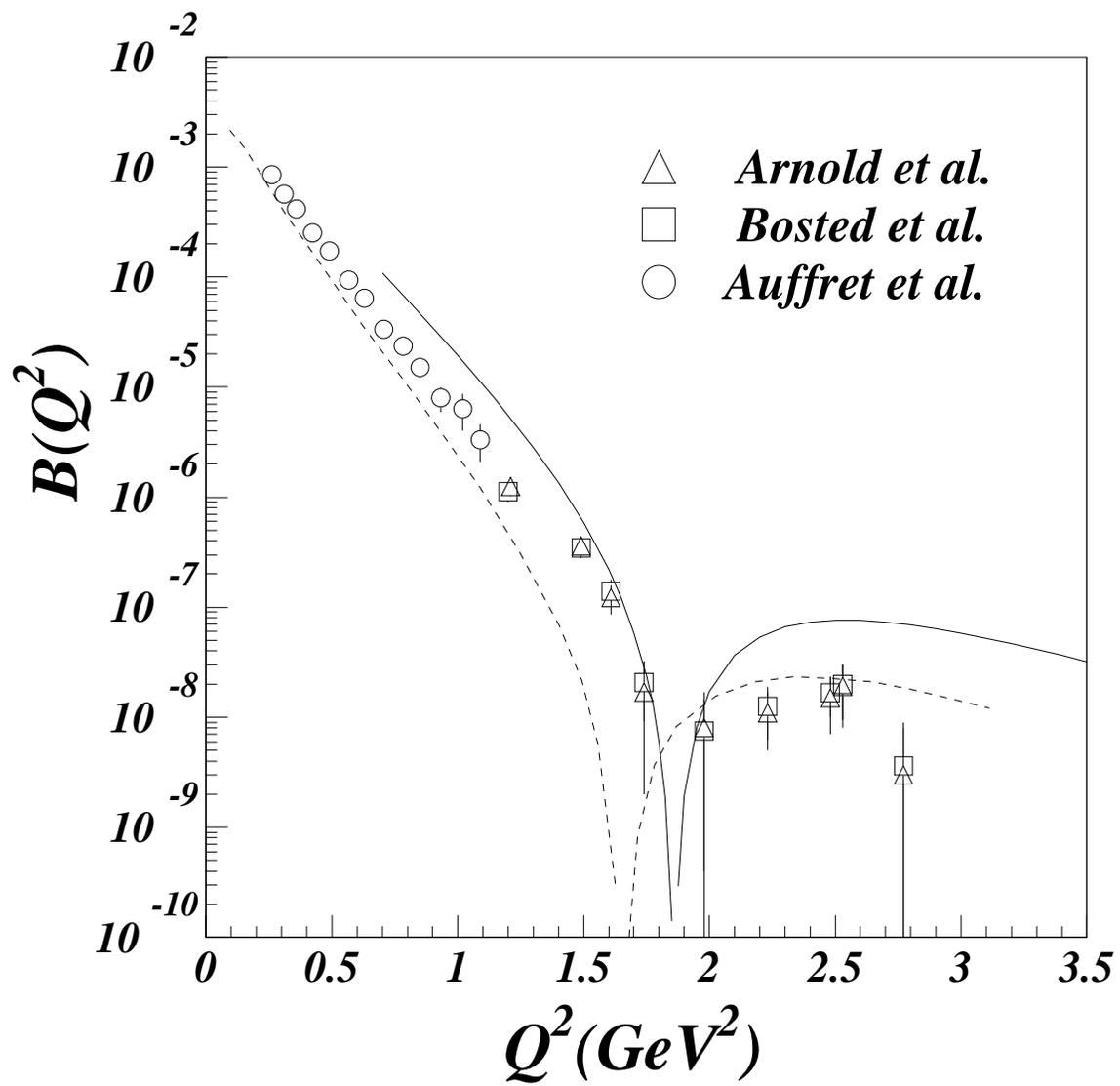
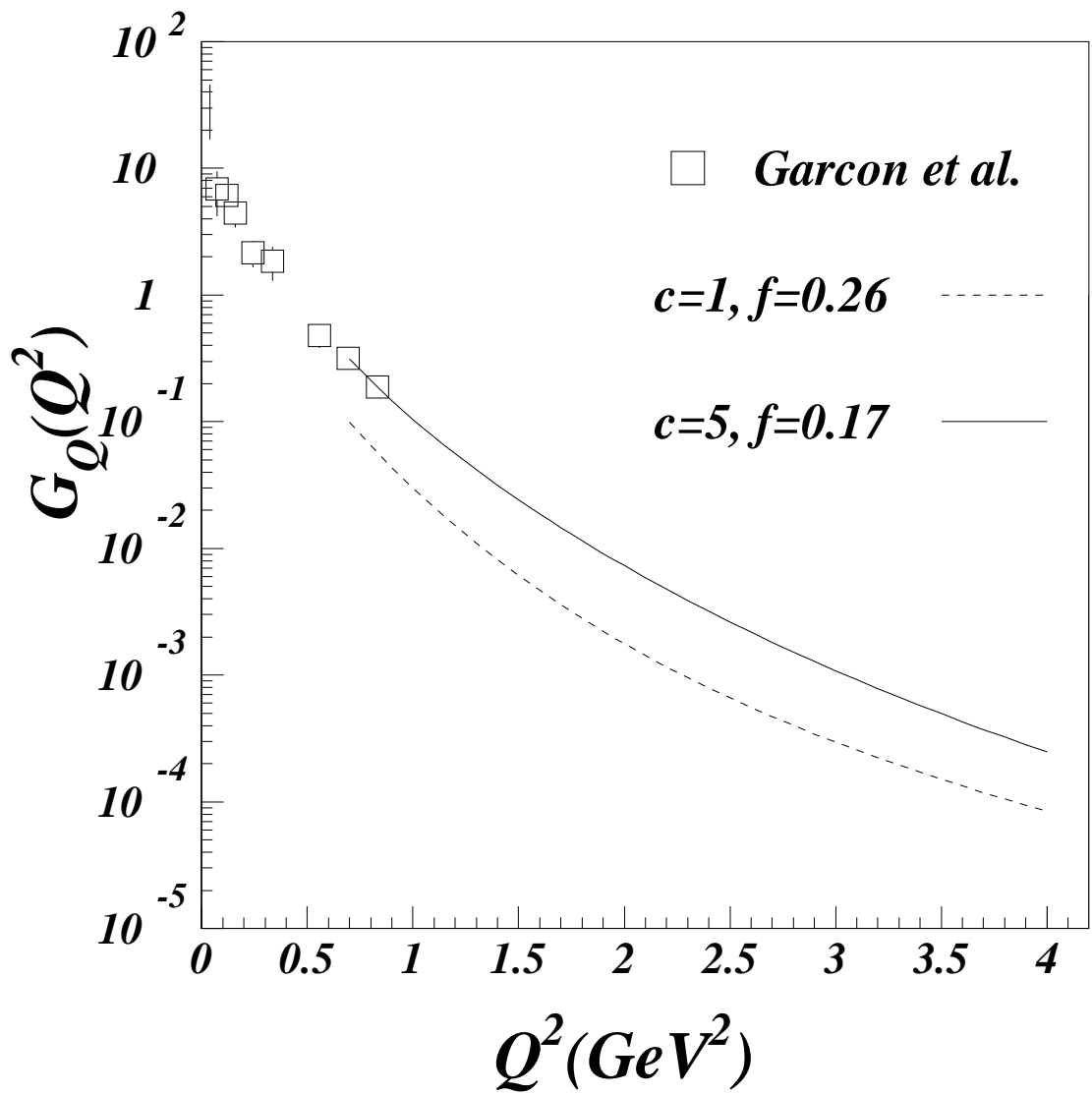
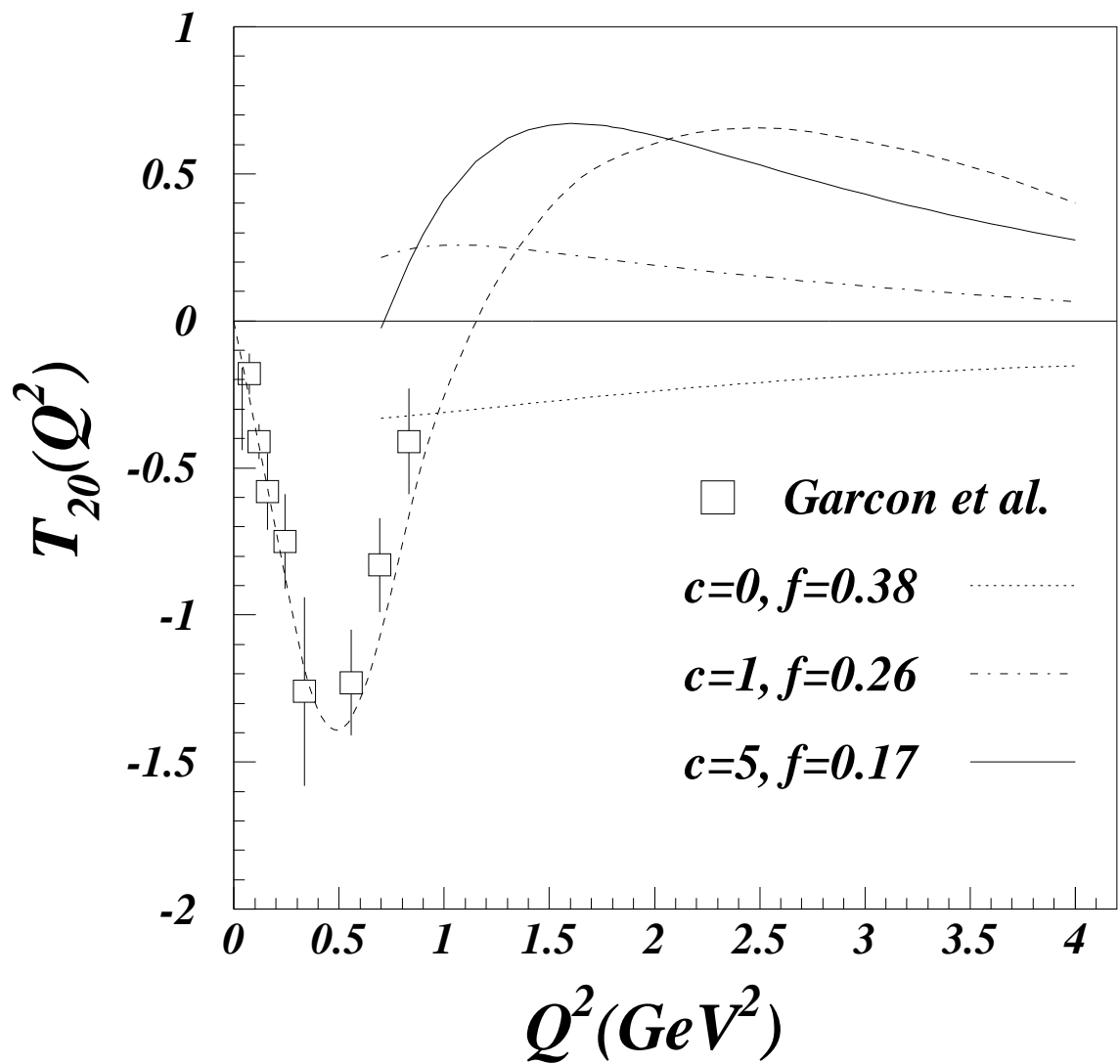
\end{document}